\documentclass[aps,pre,preprint,floatfix]{revtex4}
\usepackage{graphicx,graphics,epsf,psfig,epsfig}
\usepackage{subfigure}

\begin{document}

\title{Geometrical model for the native-state folds of proteins}

\author{Trinh X. Hoang $^1$, Antonio Trovato $^{2}$, Flavio
Seno$^{2}$, Jayanth R. Banavar $^3$, and Amos Maritan $^{2}$}
\affiliation{(1) Institute of Physics and Electronics, VAST, 10 Dao
Tan, Hanoi, Viet Nam\\ (2) INFM and Dipartimento di Fisica
`G. Galilei', Universit\`a di Padova, Via Marzolo 8, 35131 Padova,
Italy\\ (3) Department of Physics, 104 Davey Lab, The Pennsylvania
State University, University Park PA 16802, USA}

\vspace{1cm}

\begin{abstract}

We recently introduced a physical model \cite{HoangPNAS,PRE} for proteins
which incorporates, in an approximate manner, several key features such as
the inherent anisotropy of a chain molecule, the geometrical and energetic
constraints placed by the hydrogen bonds and sterics, and the role played
by hydrophobicity. Within this framework, marginally compact conformations
resembling the native state folds of proteins emerge as broad competing
minima in the free energy landscape even for a homopolymer. Here we show
how the introduction of sequence heterogeneity using a simple scheme of
just two types of amino acids, hydrophobic (H) and polar (P), and sequence
design allows a selected putative native fold to become the free energy
minimum at low temperature. The folding transition exhibits thermodynamic
cooperativity, if one neglects the degeneracy between two different low
energy conformations sharing the same fold topology.

\end{abstract}

\maketitle

\newcounter{ctr}
\setcounter{ctr}{1}

\section{Introduction}

Proteins are well-tailored chain molecules employed by life to store
and replicate information, to carry out a dizzying array of
functionalities and to provide a molecular basis for natural
selection. A protein molecule is a large and complex physical system
with many atoms. In addition, the water molecules surrounding the
protein play a crucial role in its behavior.  At the microscopic
level, the laws of quantum mechanics can be used to deduce the
interactions but the number of degrees of freedom are far too many for
the system to be studied in all its detail.  When one attempts to look
at the problem in a coarse-grained manner\cite{bmedit} with what one
hopes are the essential degrees of freedom, it is very hard to
determine what the effective potential energies of interaction are.
This situation makes the protein problem particularly daunting and no
solution has yet been found. Nevertheless, proteins fold into a
limited number\cite{ChFin90,Chothia1} of evolutionarily conserved
structures\cite{Denton,Chothia2}. The same fold is able to house many
different sequences which have that conformation as their native state
and is also employed by nature to perform different biological
functions, pointing towards the existence of an underlying simplicity
and of a limited number of key principles at work in proteins.

We have recently shown that a simple model which encapsulates a few
general attributes common to all polypeptide chains, such as steric
constraints\cite{Rama,LINUS,Baldwin}, hydrogen
bonding\cite{Pauling1,Pauling2,Eisen} and
hydrophobicity\cite{Kauzmann}, gives rise to the emergent free energy
landscape of globular proteins \cite{HoangPNAS,PRE}.  The relatively few
minima in the resulting landscape correspond to distinct putative
marginally-compact native-state structures of proteins, which are
tertiary assemblies of helices, hairpins and planar sheets. A superior
fit\cite{Frustration,Brenner} of a given protein or sequence of amino
acids to one of these pre-determined folds dictates the choice of the
topology of its native-state structure. Instead of each sequence
shaping its own free energy landscape, we find that the overarching
principles of geometry and symmetry determine the menu of possible
folds that the sequence can choose from.

Sequence design would favor the appropriate native state
structure over the other putative ground states leading to a energy
landscape conducive for rapid and reproducible folding of that
particular protein. Nature has a choice of 20 amino acids for the
design of protein sequences. A pre-sculpted landscape greatly
facilitates the design process. Indeed we will show in detail that,
within our model, a crude design scheme, which takes into account the
hydrophobic (propensity to be buried) and polar (desire to be exposed
to the water) character of the amino acids, is sufficient to carry out
a successful design of sequences with one of the 
structures shown in Fig. \ref{Fig1}.

\section{Model and Methods}

We model a protein as a chain of {\em identical} amino acids,
represented by their $C^{\alpha}$ atoms, lying along the axis of a
self-avoiding flexible tube.  The preferential parallel placement of nearby
tube segments approximately mimics the effects of
the anisotropic interaction of hydrogen bonds,
while the space needed for the clash-free packing of side chains is
approximately captured by the non-zero tube thickness\cite{Tubone,RMP,Thick}.
A tube description places constraints on the radii of circles drawn through
both local and non-local triplets of  $C^{\alpha}$ positions of a
protein native structure\cite{RMP,BMMT02}.

Unlike unconstrained matter for which pairwise interactions suffice,
for a chain molecule, it is necessary to define the context of the
object that is part of the chain.  This is most easily carried out by
defining a local Cartesian coordinate system whose three axes are
defined by the tangent to the chain at that point, the normal, and the
binormal which is perpendicular to both the other two vectors. A
study\cite{HoangPNAS,PRE} of the experimentally determined native
state structures of proteins from the Protein Data Bank reveals that
there are clear amino acid aspecific geometrical constraints on the
relative orientation of the local coordinate systems due to sterics
and also associated with amino acids which form hydrogen bonds with
each other. Similar geometrical constraints had already been
introduced in off-lattice polymer models \cite{Jeff,Jesper} in order
to model hydrogen bond formation.

The geometrical constraints associated with the formation of hydrogen
bonds and with the tube description within the $C^{\alpha}$
representation of our model are described in detail
elsewhere\cite{HoangPNAS,PRE}. In our representation of the protein
backbone, local hydrogen bonds form between $C^{\alpha}$ atoms
separated by three along the sequence with an energy defined to be
$-1$ unit, whereas non-local hydrogen bonds are those that form
between $C^{\alpha}$ atoms separated by more than $4$ along the
sequence with an energy of $-0.7$.  This energy difference is based on
experimental findings that the local bonds provide more stability to a
protein than do the non-local hydrogen
bonds\cite{Sosnick}. Cooperativity effects\cite{Liwo,Fain} are taken
into account by adding an energy of $-0.3$ units when consecutive
hydrogen bonds along the sequence are formed.  There are two other
ingredients in the model: a local bending penalty $e_R$ which is
related to the steric hindrance of the amino acid side chains and a
pair-wise interaction $e_W$ of the standard type mediated by the
water\cite{Kauzmann}. Note that whereas the geometrical constraints
associated with the tube and hydrogen bonds are representative of the
typical {\em aspecific} behavior of the interacting amino acids, the
latter properties clearly depend on the {\em specific} amino acids
involved in the interaction.

Monte Carlo simulations have been carried out with pivot and crankshaft moves
commonly used in stochastic chain dynamics \cite{Sokal}. A Metropolis
procedure is employed with a thermal weight $\exp\left(-E/T\right)$,
where $E$ is the energy of the conformation and $T$ is the effective
temperature.

\section{Results and discussion}

\begin{figure}
\centering \includegraphics[width=8.0cm]{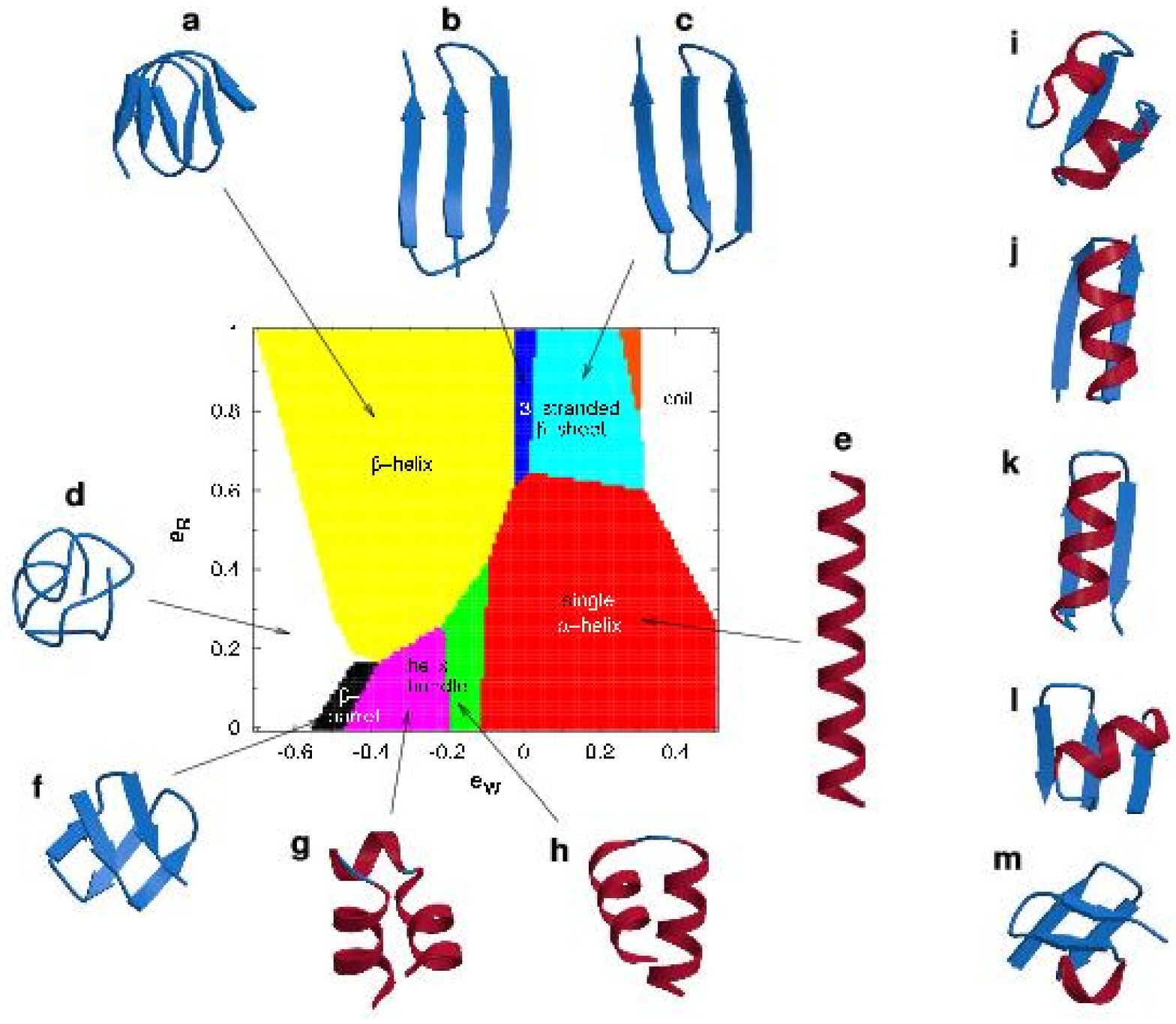}
\caption{Phase diagram of ground state conformations.  The ground
state conformations were obtained by means of Monte-Carlo simulations
of chains of 24 $C^{\alpha}$ atoms. $e_R$ and $e_W$ denote the bending
energy penalty and the solvent mediated interaction energy
respectively. Over $600$ distinct local minima were obtained in our
simulations in different parts of parameter space starting from a
randomly generated initial conformation. The temperature is set
initially at a high value and then decreased gradually to zero.  (a),
(b), (c), (e), (f), (g), (h) are the Molscript representation of the
ground state conformations which are found in different parts of the
parameter space as indicated by the arrows. The helices and strands
are assigned when local or non-local hydrogen bonds are formed
according to the rules employed within our
model\cite{HoangPNAS}. Conformations (i), (j), (k), (l), (m) are
competitive local minima. In the orange phase, the ground state is a
2-stranded $\beta$-hairpin (not shown). Two distinct topologies of a
3-stranded $\beta$-sheet (dark and light blue phases) are found
corresponding to conformations shown in conformations (b) and (c)
respectively.  The white region in the left of the phase diagram has
large attractive values of $e_W$ and the ground state conformations
are compact globular structures with little amount of secondary
structures.  At even lower values of $e_W$, the ground states exhibit
a crystalline order induced by hard sphere packing
considerations\cite{Karplus} and not by hydrogen bonding (conformation
(d)).}
\label{Fig1}
\end{figure}

Figure \ref{Fig1} shows the ground state phase diagram obtained from
Monte-Carlo computer simulations using the simulated annealing
technique\cite{Anneal}, along with the corresponding conformations,
for a $24$ beads homopolymer \cite{HoangPNAS,PRE}. The solvent-mediated
energy, $e_W$, and the local bending penalty, $e_R$, are measured in
units of the local hydrogen bond energy.  When $e_W$ is sufficiently
repulsive (hydrophilic) (and $e_R > 0.3$ in the phase diagram), one
obtains a swollen phase with very few contacts between the
$C^{\alpha}$ atoms. When $e_W$ is sufficiently attractive, one finds a
very compact, globular phase with featureless ground states with a
high number of contacts.

Between these two phases (and in the vicinity of the swollen phase), a
marginally compact phase emerges (the interactions barely stabilize
the ordered phase) with distinct structures including a single helix,
a bundle of two helices, a helix formed by $\beta$-strands, a
$\beta$-hairpin, three-stranded $\beta$-sheets with two distinct
topologies and a $\beta$-barrel like conformation. These
structures are the stable ground states in different parts of the
phase diagram.  Furthermore, other conformations, closely resembling
distinct super-secondary arrangements observed in
proteins\cite{ChFin90}, also shown in Fig. \ref{Fig1}, are found to be
competitive local minima, whose stability can be enhanced, as we will
see, by sequence design after heterogeneity is introduced by means of,
for example, non-uniform values of curvature energy penalties for
single amino acids and hydrophobic interactions for amino acid pairs.
Note that while there is a remarkable similarity between the
structures that we obtain and protein folds, our simplified
coarse-grained model is not as accurate as an all-atom representation
of the poly-peptide chain in capturing features such as the packing of
amino acid side chains.

The common belief in the field of proteins is that given a
sequence of amino acids, with all the attendant details of the side
chains and the surrounding water, one obtains a funnel-like landscape
with the minimum corresponding to its native state structure. Each
protein is characterized by its own landscape.  In this scenario, the
protein sequence is all-important and the protein folding problem,
besides becoming tremendously complex, needs to be attacked on a
protein-by-protein basis.

In contrast, our model calculations show that the large number of
common attributes of globular proteins\cite{RMP,Bernal} reflect
a deeper underlying unity in their behavior. At odds with conventional
belief, the gross features of the energy landscape of proteins result
from the amino acid aspecific common features of all proteins, as is
clearly established by the fact that different putative native
structures are found to be competing minima for the same homopolymeric
chain.

\begin{figure}
\centering \includegraphics[width=8.0cm]{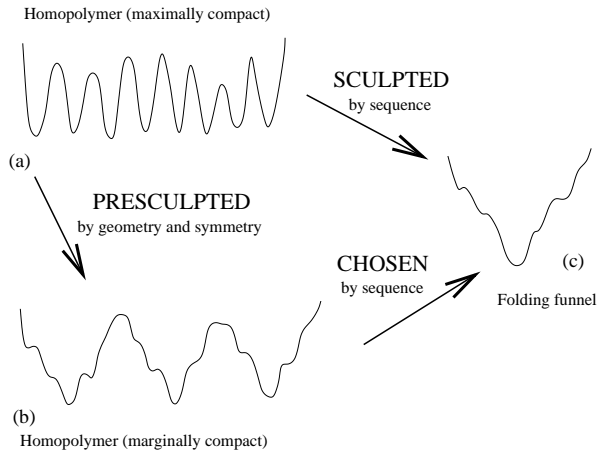}
\caption{Simplified one dimensional sketches of energy landscape. The
quantity plotted on the horizontal axis schematically represents a
distance between different conformations in the phase space and the
barriers in the plots indicate the energy needed by the chain in order
to travel between two neighboring local minima. (a) Rugged energy
landscape for a homopolymer chain with an attractive potential
promoting compaction as, e.g., in a string and beads model. There are
many distinct maximally compact ground state conformations with
roughly the same energy, separated by high energy barriers (the
degeneracy of ground state energies would be exact in the case of both
lattice models and off-lattice models with discontinuous square-well
potentials). (b) Presculpted energy landscape for a homopolymer chain
in the marginally compact phase. The number of minima is greatly
reduced and the width of their basin increased by the introduction of
geometrical constraints. (c) Funnel energy landscape for a protein
sequence. As folding proceeds from the top to the bottom of the
funnel, its width, a measure of the entropy of the chain, decreases
cooperatively with the energy gain. Such a distinctive feature,
crucial for fast and reproducible folding, arises from careful
sequence design in models whose homopolymer energy landscape is
similar to (a). In contrast, funnel-like properties already result
from considerations of geometry and symmetry in the marginally compact
phase (b), thereby making the goals of the design procedure the
relatively easy task of stabilization of one of the pre-sculpted
funnels followed by the more refined task of fine-tuning the putative
interactions of the protein with other proteins and ligands.}
\label{Fig3}
\end{figure}

The protein energy landscape is {\em (pre)sculpted} by general
considerations of geometry and symmetry (Fig. \ref{Fig3}), which are
the essential ingredients in our model. Our unified framework suggests
that the protein energy landscape ought to have around a thousand of
broad minima corresponding to putative native state structures.  The
key point is that for each of these minima the desirable funnel-like
behavior is already achieved at the homopolymer level {\em in the
marginally compact part of the phase diagram}. The self-tuning of two
key length scales, the thickness of the tube and the interaction
range, to be comparable to each other and the interplay of the three
energy scales, hydrophobic, hydrogen bond, and bending energy, in such
a way as to stabilize marginally compact structures, also provide the
close cooperation between energy gain and entropy loss needed for the
sculpting of a funneled energy landscape. At the same time, relatively
small changes in the parameters $e_W$ and $e_R$ lead to significant
differences in the emergent ground state structure, underscoring the
sensitive role played by chemical heterogeneity in selecting from the
menu of native state folds.

The introduction of sequence heterogeneity at the level of
differantiating hydropohobic (H) and polar (P) residues and a crude
designe scheme based on a common sense choice of the hydrophobicity
profile for the sequence suffices to ensure that the designed sequence
would fold into a desired fold selected from the predetermined
menu. For example, the $\beta$-$\alpha$-$\beta$ motif shown as (j) in
Fig. \ref{Fig1} (which is a local energy minimum for a homopolymer)
can be stabilized into a global energy minimum for the sequence
HPHHHPPPPHHPPHHPPPPHHHPP, with $e_W=-0.4$ for HH contacts and $e_W=0$
for other contacts, and $e_R=0.3$ for all residues.

\begin{figure}
\hfill \subfigure[]{\includegraphics[height=5.5cm]{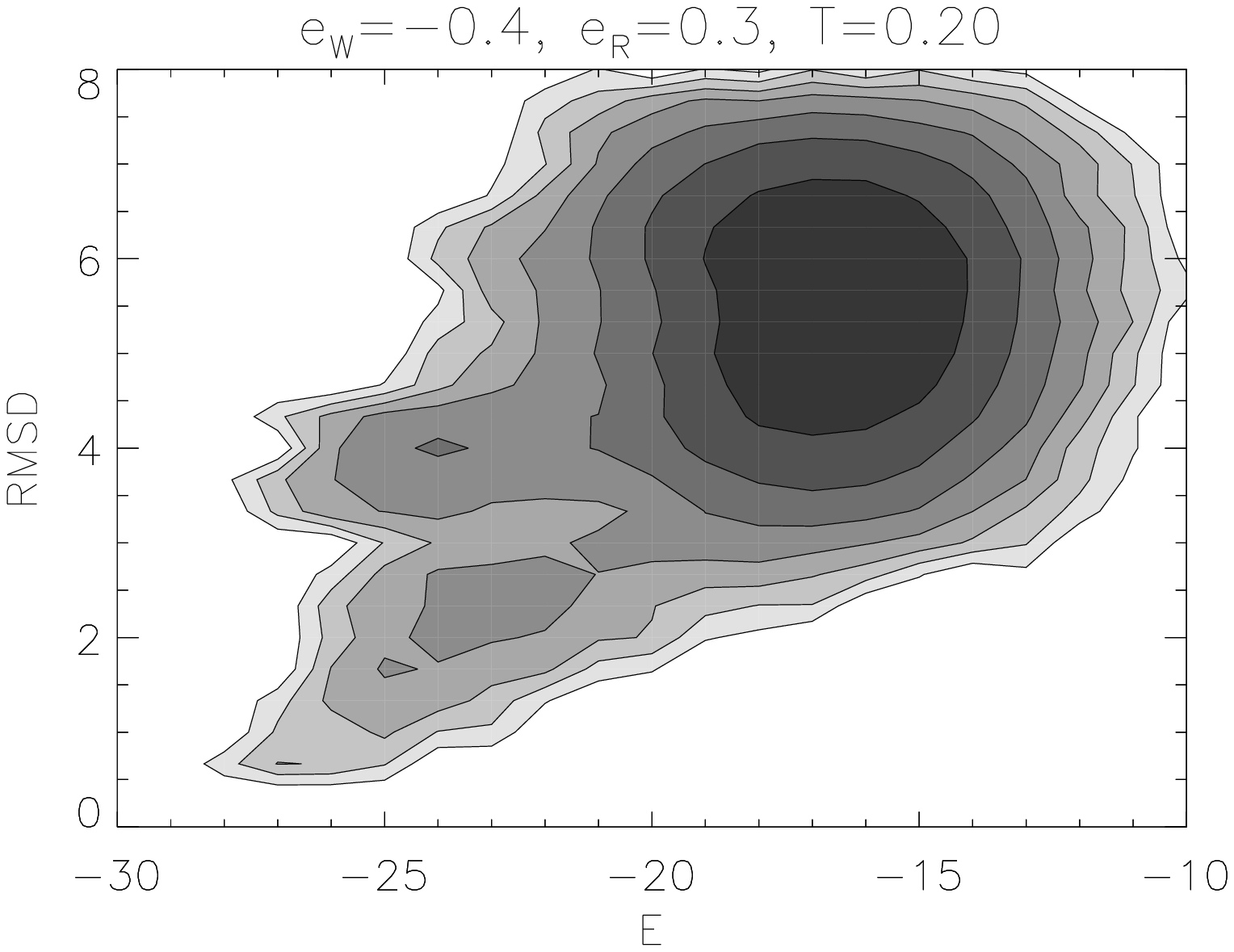}} \hfill
\subfigure[]{\includegraphics[height=5.5cm]{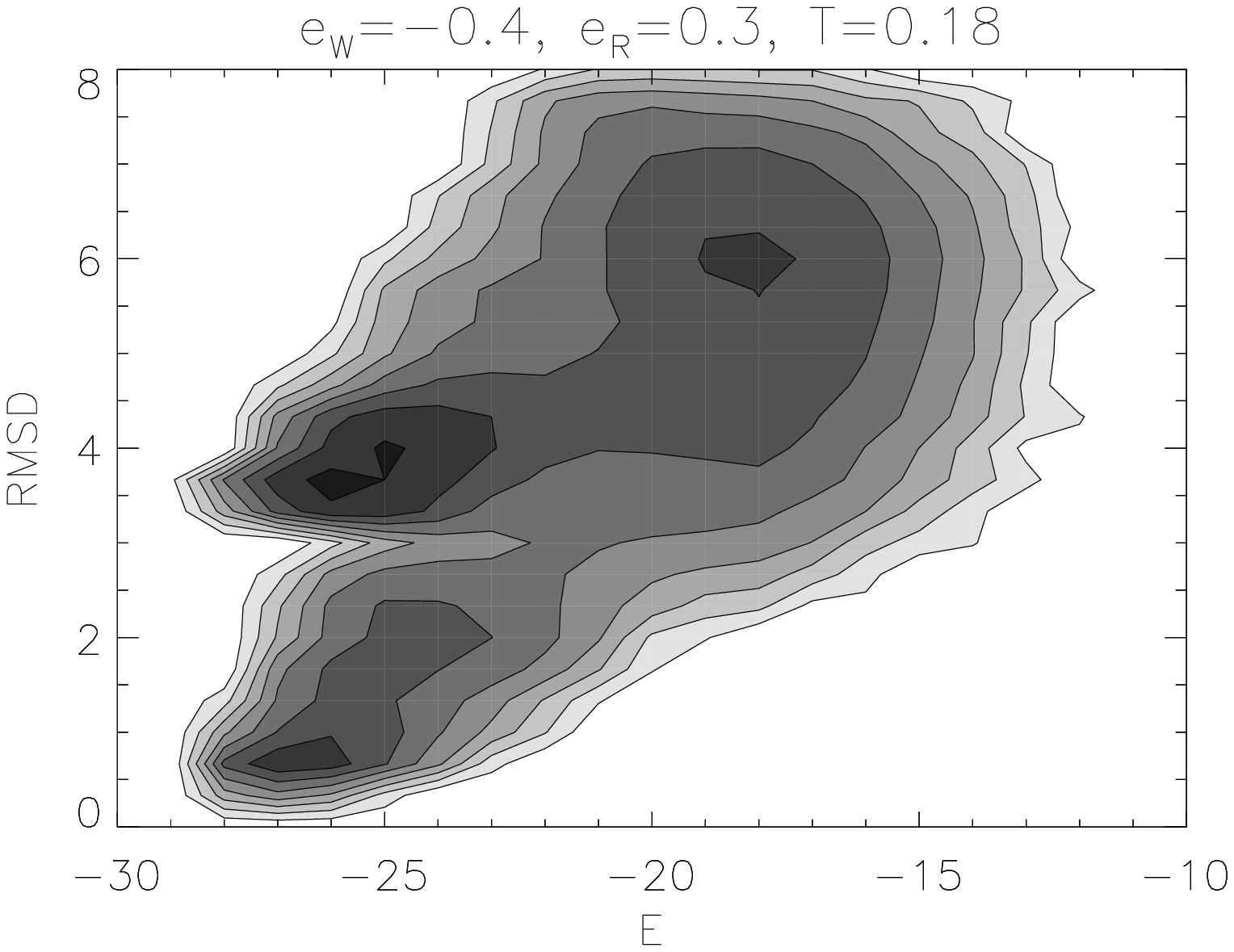}} \hfill \\
\hfill \subfigure[]{\includegraphics[height=3.5cm]{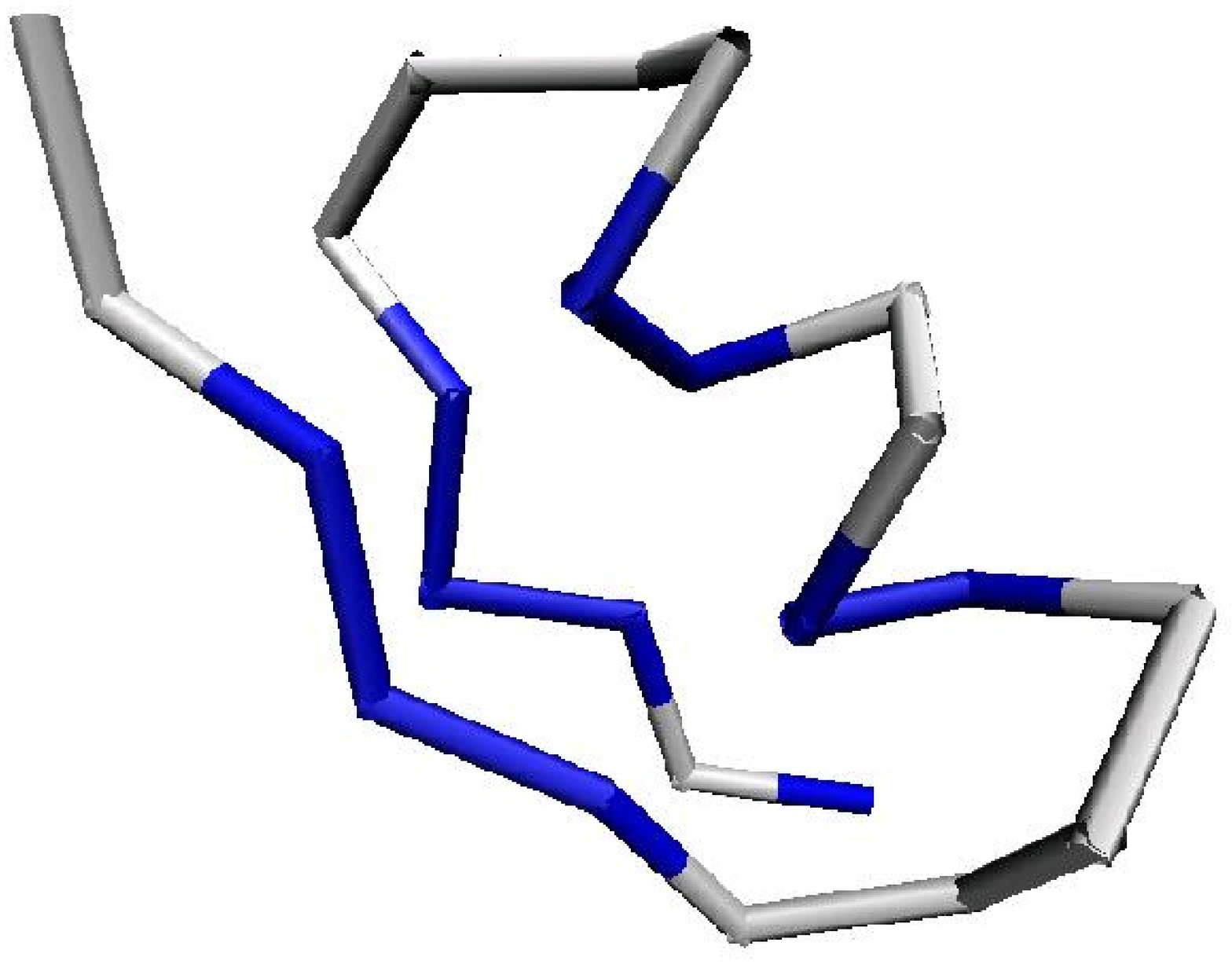}} \hfill
\subfigure[]{\includegraphics[height=3.5cm]{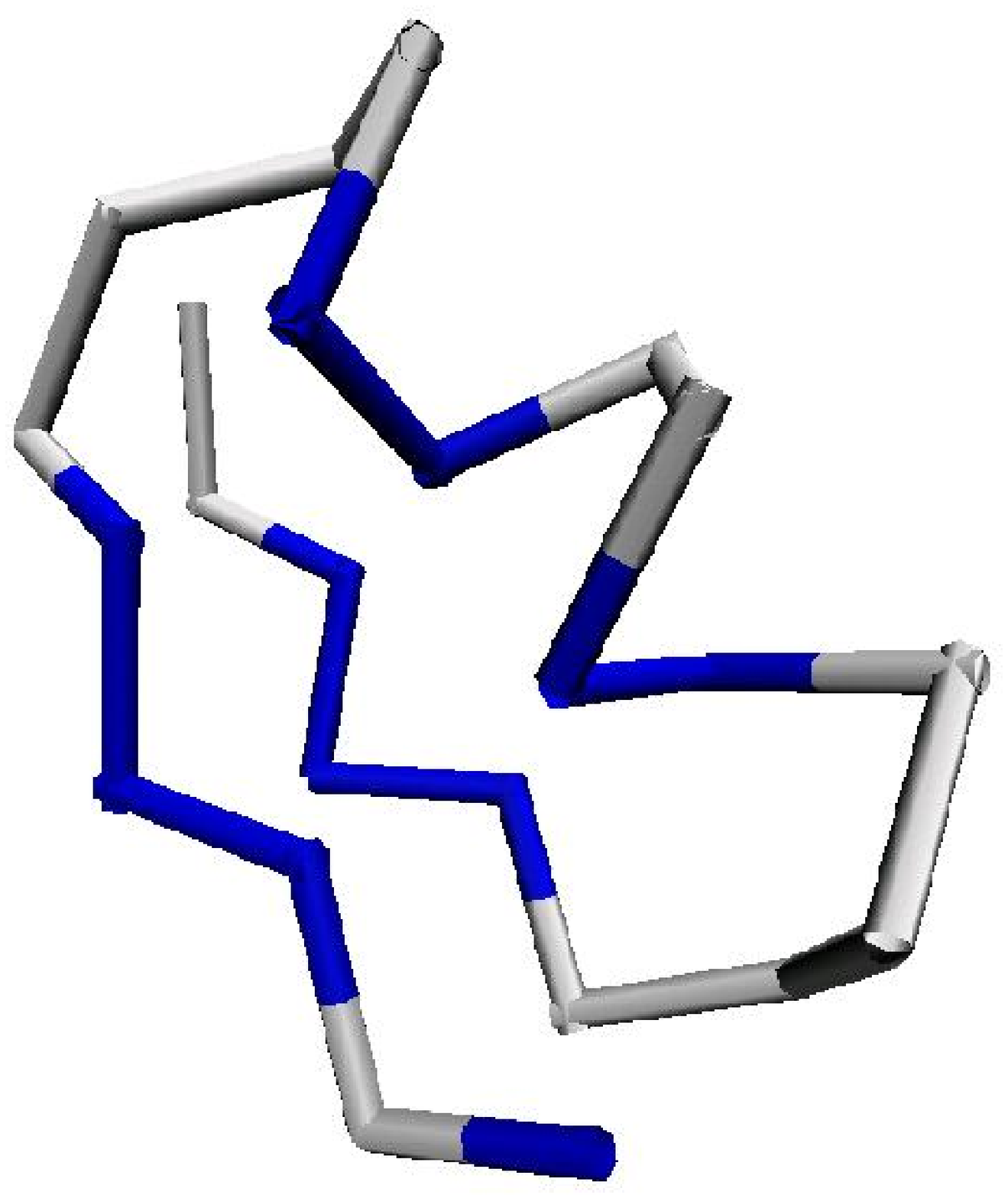}} \hfill \\
\caption{Contour plots of the effective free energy (a) at high
temperature ($T = 0.20$) and (b) at the folding transition temperature
$T_f=0.18$ for the HPHHHPPPPHHPPHHPPPPHHHPP 24-residue chain, with
$e_W=-0.4$ for HH interacions, $e_W=0$ for other interactions, and
$e_R=0.3$. The effective free energy, defined as $F(E,RMSD)=-\ln
N(E,RMSD)$, is obtained as a function of the energy $E$ and the root
mean squared deviation $RMSD$ from the `reference' conformation shown in 
Fig. \ref{Fig1}(j), from the histogram $N(E,RMSD)$ collected in
equilibrium Monte-Carlo simulations at constant temperature. The
spacing between consecutive levels in each contour plot is $1$ and
corresponds to a free energy difference of $k_B\tilde{T}$, where
$\tilde{T}$ is the temperature in physical units. The darker the
color, the lower the free energy value.  There is just one free energy
minimum corresponding to the denatured state at a temperature higher
than the folding transition temperature (Panel (a)) whereas one can
discern the existence of three distinct minima at the folding
transition temperature (Panel (b)). The two conformations
corresponding to the two low energy minima are shown in Panel (c)
(higher RMSD) and (d) (lower RMSD). Hydrophobic (polar) residues are
shown in dark (light) grey} \label{Fig4}
\end{figure}

We have studied the thermodynamic properties of the folding transition
for this case. The contour plots at different temperatures (above and
at the folding transition temperature) of the effective free energy
are shown in Fig. \ref{Fig4}(a) and \ref{Fig4}(b) as a function of the
total energy of the chain and its root mean squared deviation (RMSD)
from the $\beta$-$\alpha$-$\beta$ conformation obtained as a local
minimum in the homopolymer case.  The folding transition temperature
was identified at $T=0.18$ from the peak of the specific heat curve
(data not shown).

The free energy landscape is not too smooth at low temperatures, but
it can be roughly be described as a three state landscape. Above the
folding transition there is one broad minimum, corresponding to the
denatured state ensemble. The two clearly distinct minima which
dominate the free energy landscape below the folding transition
correspond to the designed conformation (in the case of the lower
RMSD) and to a different conformation (with higher RMSD). Both
structures share the same number of HH contacts as well as local,
non-local, and cooperative hydrogen bonds, differing just in the way
the two strands are connected with the helix (see Figure \ref{Fig4}(c)
and \ref{Fig4}(d)). Note that both minima are consistently
characterized by roughly the same energy.

This degeneracy is due to the lack of detail of the model and to the
crudeness of the design scheme utilized here, since the hydrophobic
core is well formed in both cases (see Fig. \ref{Fig4}(c) and
\ref{Fig4}(d)). Indeed, if one groups both conformations within the
same fold classification, this is a clear example of how, in the
presence of a presculpted energy landscape, it is relatively easy to
design a sequence with the ability to fold cooperatively into a given
fold, as real small globular proteins do.

\section{Conclusions}

In summary, within a simple, yet realistic, framework, we have shown
\cite{HoangPNAS,PRE} that protein native-state structures can arise from
considerations of symmetry and geometry associated with the
polypeptide chain. The sculpting of the free energy landscape with
relatively few broad minima is consistent with the fact that proteins
can be designed to enable rapid folding to their native states. Here
we have shown that by introducing heterogeneity within the simplest
hydrophobic-polar scheme, it is straightforward to design a sequence
that is able to fold cooperatively into one of the presculpted minima
in the energy landscape.

\vspace{1cm}

\noindent {\bf Acknowledgements} This work was supported by PRIN 2003,
FISR 2001, NASA, NSF IGERT grant DGE-9987589, NSF MRSEC, and VNSC.

\newpage

\newpage

\newpage

\end{document}